\documentclass[aps,pre,a4paper,preprint,showpacs,showkeys,superscriptaddress]{revtex4-1}%
\usepackage{graphicx}
\usepackage{amssymb,amsmath,amsthm,amsfonts}
\usepackage{subfigure}
\usepackage{xcolor}
\usepackage{amsmath}
\usepackage{amsfonts}
\usepackage{amssymb}%
\begin{document}
\title{Scaling Analysis of Random Walks with Persistence Lengths: Application to
Self-Avoiding Walks}
\author{\firstname{C.} R. F. \surname{Granzotti}}
\email{c.roberto.fg@usp.br}
\author{\firstname{A.} S. \surname{Martinez}}
\email{asmartinez@usp.br}
\altaffiliation{National Institute of Science and Technology in Complex Systems (INCT-SC)}
\affiliation{Faculdade de Filosofia, Ci\^encias e Letras de Ribeir\~ao Preto
(FFCLRP),Universidade de S\~ao Paulo (USP), Avenida Bandeirantes 3900, CEP
14040-901, Ribeir\~ao Preto, S\~ao Paulo, Brazil.}
\author{\firstname{M.} A. A. da \surname{Silva}}
\email{maasilva@fcfrp.usp.br}
\affiliation{Faculdade de Ci\^{e}ncias Farmac\^{e}uticas de Ribeir\~{a}o Preto (FCFRP),
Universidade de S\~{a}o Paulo (USP), Avenida Bandeirantes 3900, CEP 14040-901,
Ribeir\~{a}o Preto, S\~{a}o Paulo, Brazil.}


\begin{abstract}
We develop an approach for performing scaling analysis of $N$-step Random
Walks (RWs). The mean square end-to-end distance, $\langle\vec{R}_{N}%
^{2}\rangle$, is written in terms of inner persistence lengths (IPLs), which
we define by the ensemble averages of dot products between the walker's
position and displacement vectors, at the $j$-th step. For RW models statistically invariant under orthogonal transformations, we analytically introduce a relation between $\langle\vec{R}_{N}^{2}\rangle$ and the persistence length, $\lambda_{N}$, which is defined as the mean end-to-end vector projection in the first step direction. For Self-Avoiding Walks (SAWs) on
2D and 3D lattices we introduce a series expansion for $\lambda_{N}$, and by
Monte Carlo simulations we find that $\lambda_{\infty}$ is equal to a
constant; the scaling corrections for $\lambda_{N}$ can be second and higher
order corrections to scaling for $\langle\vec{R}_{N}^{2}\rangle$. Building
SAWs with typically one hundred steps, we estimate the exponents
$\nu_{0}$ and $\Delta_{1}$ from the IPL behavior as function of $j$. The
obtained results are in excellent agreement with those in the literature.
This shows that only an ensemble of paths with the same length is
sufficient for determining the scaling behavior of $\langle\vec{R}_{N}%
^{2}\rangle$, being that the whole information needed is contained in the
inner part of the paths.

\end{abstract}
\volumeyear{year}
\volumenumber{number}
\issuenumber{number}
\eid{identifier}
\date{\today}
\maketitle

\section{INTRODUCTION}

Random Walk (RW) models are ubiquitous in the literature with applications in several areas, such as Physics~\cite{N_Stanley_2001}, Biology~\cite{JRSI_Codling_2008} and Economy \cite{Book_IEC_1999}. The simplest case is the walker displacement in a sequence of independent random steps, namely ordinary RW~\cite{Book_FSRW_2013}. One may also obtain random paths on a geometrical space with distinct displacement schemes, leading to other RW models. A fundamental importance of these models lies in the fact that many real phenomena can be mapped or directly represented by
paths traversed by walkers in some geometrical space, e.g., a single-strand DNA~\cite{JCP_Rechendorff_2009} and magnetic systems~\cite{PRB_Heiko_1998}. An example is the Self-Avoiding Walk (SAW) defined by a walker forming a random path that never intersects itself; standard SAWs are performed on regular lattices, where the walker steps to nearest-neighbor sites and does not visit a site more than once \cite{Book_Madras_2012}. 

Because of non-overlapping paths, the SAW model plays a central role in Polymer Physics \cite{Book_PP_2003} by capturing the excluded volume effect in a dilute solution under good solvent condition or at high temperatures~\cite{PRB_Stanley_1987}. The SAW model is also well known in statistical physics context because of its equivalence with the $n$-vector model with $n\rightarrow0$, as de Gennes first pointed out \cite{Book_DeGennes_1979}. From this equivalence, with arguments
of renormalization and field theories, one expects the following series expansion for the mean square end-to-end distance
\cite{JPA_Janse_2009,Belorec_1997}:
\begin{multline}
\langle\vec{R}_{N}^{2}\rangle_{N}=a_{0}N^{2\nu_{0}}
(1+\frac{a_{1}}
{N}+\frac{a_{2}}{N^{2}}+\cdots\frac{b_{1}}{N^{\Delta_{1}}}+ \\
+\frac{b_{2}}{N^{\Delta_{1}+1}}+\cdots  \frac{c_{1}}{N^{\Delta_{2}}%
}+\frac{c_{2}}{N^{\Delta_{2}+1}}+\cdots), 
\label{assintotic}
\end{multline}
where $\nu_{0}$ is the leading exponent. The terms proportional to $N^{-i}$ with $i=1,2,\cdots$, are analytical corrections, and the terms proportional to $N^{-(j+\Delta_{i})}$ with non-integer exponents $\Delta_{i}<\Delta_{i+1}$ and $j=0,1,2\cdots$, are the non-analytical corrections to scaling. The leading and corrections to scaling exponents are universal. The indexed brackets $\langle.\rangle_{N}$ refers to the $N$-step RW ensemble average, and from now on, unless strictly necessary, we omit the
index $N$. Numerical estimates of exponents $\nu_{0}$ and $\Delta_{1}$ are based on either exact counting techniques \cite{PRL_Fricke_2014,JSM_Schram_2011}, or in Monte Carlo (MC) simulation methods~\cite{JSP_Madras_1988,JSP_HPS_2011}, through the sampling of $\langle\vec{R}_{N}^{2}\rangle$ \cite{PRL_Clisby_2010,JSP_Clisby_2010}. 

Obtaining such estimates for $\nu_{0}$ and $\Delta_{1}$, especially for 3D SAW, is a challenge from several points of view. The exponential growth of the number of possible $N$-step paths $c_{N} \approx \mu^{N}N^{\gamma-1}$, where $\mu$ is the connectivity constant and $\gamma>1$, imposes a limit to exact counting. To the best of our knowledge, the maximum values obtained are $N=79$~\cite{Arx_Jensen_2013} and $N=36$~\cite{JSM_Schram_2011} for SAWs on 2D and 3D square lattices, respectively. Concerning Monte Carlo simulations, there exist an appeal to find $\nu_{0}$ and $\Delta_{1}$ using very long paths. Obtaining high quality Monte Carlo data for such path lengths is an extremely difficult task for the SAW model. The variable length algorithms suffer from attrition problems, namely barriers that prevent paths to grow, while the fixed length algorithms suffer from the decreasing of acceptance rate to generate a new non-self-intersecting path, according to the increase of the (fixed) path length~\cite{NPB_Sokal_1996}.

Numerical drawbacks also take place when one studies other conformational quantities. An example is the persistence length, $\lambda_{N}$, defined as the mean end-to-end vector projection in a fixed direction along the first step~\cite{Book_Cantor_1980,PNAS_Flory_1973}, as $N\rightarrow\infty$~\cite{Note1}. Defining the end-to-end vector as $\vec{R}_{N}=\sum_{j=1}^{N}\vec{u}_{j}$, where $\vec{u}_{j}$ is the walker displacement at the $j$-th step, the persistence length can be expressed by $\lambda_{N}=\langle\vec{R}_{N}\cdot\vec{u} _{1}\rangle/|\vec{u}_{1}|=\sum_{j=1}^{N}\langle\vec{u}_{1}\cdot\vec{u}
_{j}\rangle/|\vec{u}_{1}|$. Numerical results of $\lambda_{N}$, for 2D-SAWs, are controversial in the literature, and for 3D, are scarce~\cite{AIP_Hagai_1990}. For 2D-SAW, Grassberger~\cite{PLA_Grassberger_1982} obtained the first estimate of $\lambda_{N}$ in the square lattice, by means of a power law $\lambda_{N}\sim N^{\theta}$, with $\theta=0.063(10)$. Since for $\theta\approx0$, it is also well fitted by $\lambda_{N}\sim\ln(N)$, as suggested by Redner and Privmann~\cite{JPA_Privman_1987}. They obtained both estimates by sampling the displacements projections along the first step direction, for all possible configurations of SAW paths with $N<24$. This weak divergence has been questioned recently by Eisenberg and Baram \cite{JPA_Eisenberg_2003}, because their MC estimates of $\langle\vec{u}_{1}\cdot\vec{u}_{j}\rangle$ show that $\lambda_{N}$ converges to a constant when $N\rightarrow\infty$. One could employ $\lambda_{N}$ in Monte Carlo~\cite{Group} and experimental characterization of certain polymers~\cite{PNAS_Sitlani_2000,CP_Dogsa_2014}, despite there exist some limitations of $\lambda_{N}$ measures such as divergence and edge effects~\cite{Group2}.

Refined results about the scaling behavior of the aforementioned conformational quantities to study universality are challenging, and have been the subject of discussion for many years~\cite{JSP_Madras_1988,JPA_Janse_2009}. As usually one does not have exact results for the SAW model, there exists an appeal for simulations of
large, sometimes very large, paths. Here, one proposes to answer two questions about a SAW: (i) What is the asymptotic limit of its persistence
length? (ii) Is there some way to find out its scaling behavior employing relatively small chains? To answer these questions, we found an approach
for performing scaling analysis of RWs, by focusing in the behavior of $\langle\vec{R}_{N}^{2}\rangle$. 

The structure of the paper is as follows: In Sec.~\ref{sec:2} we present the analytical results by defining the inner persistence length and their relation with $\langle\vec{R}_{N}^{2}\rangle$ and $\lambda_{N}$, for RW models statistically invariant under orthogonal transformations. In Sec.~\ref{sec:3} we provide a series expansion for $\lambda_{N}$ and obtain the scaling behavior of 2D and 3D-SAW models with Monte Carlo simulations; we also obtain reliable estimates of the exponents $\nu_{0}$ and $\Delta_{1}$ and discuss the contribution of $\lambda_{N}$ to $\langle\vec{R}_{N}^{2}\rangle$ behavior. In Sec.~\ref{sec:4} we give concluding remarks.

\section{INNER PERSISTENCE LENGTH AND ANALYTICAL RESULTS}
\label{sec:2}

We define the inner persistence length (IPL) for an $N$-step RW, by the average dot product:
$\mathcal{I}_{j}\equiv\langle\vec{R}_{j}\cdot\vec{u}_{j}\rangle$. To relate
$\langle\vec{R}_{N}^{2}\rangle$ to $\mathcal{I}_{j}$, and $\mathcal{I}_{N}$ to
$\lambda_{N}$, we write the square distance at the $j$-th step for an $N$-step
RW as: $\vec{R}_{j}^{2}=\vec{R}_{j-1}^{2}+2\vec{R}_{j}\cdot\vec{u}_{j}
-u_{j}^{2}$. Adding up $\vec{R}_{j}^{2}$, we have $\sum_{j=1}^{k}\vec{R}
_{j}^{2}=\sum_{j=1}^{k}\vec{R}_{j-1}^{2}+\sum_{j=1}^{k}2\vec{R}_{j}\cdot
\vec{u}_{j}-\sum_{j=1}^{k}|\vec{u}_{j}|^{2}$, where $\vec{R}_{0}=\vec{0}$
leads to $\sum_{j=1}^{k}\vec{R}_{j-1}^{2}=\sum_{j=1}^{k-1}\vec{R}_{j}^{2}$.
Thus, considering $|\vec{u}_{j}|=1$, we write the average $\langle\vec{R}
_{k}^{2}\rangle=2\sum_{j=1}^{k}\mathcal{I}_{j}-k$. In particular for $k=N$,
the mean square end-to-end distance is
\begin{equation}
\langle\vec{R}_{N}^{2}\rangle=2\sum_{j=1}^{N}\mathcal{I}_{j}-N.
\label{observ_quad2}
\end{equation}

Now, consider a generic class of RWs, where ensembles of $N$-step walks obey
the following invariance property: the probability distributions, of each step
$\vec{u}_{i}$, $i=1,2,...,N$, which compose a path, is invariant under
orthogonal transformations. With this, we exclude walks like the tourist model~\cite{PRE_Lima_2001_1}, where the medium disorder
\cite{Group3} breaks down such invariance
symmetries. Particularly, one considers an ensemble of $N$-step RWs obeying
the mentioned probabilistic symmetry, under a specific orthogonal
transformation $T$ given by $\vec{u}_{i}\overset{T}{\rightarrow}\vec
{u}_{N-i+1}^{\prime}$; the prime denotes the displacement vectors in the transformed
reference frame, and $\vec{u}_{i}^{\prime}=-\vec{u}_{N-i+1}$, with
$i=1,2,...N$. Notice that $\vec{u}_{i}^{\prime}\in\{\vec{u}_{1},\vec{u}
_{2}...\vec{u}_{N}\}$, where $\{.\}$ represents the complete ensemble of
paths. This symmetry operation can be achieved by a translation followed by
inversion of all displacement vectors. In other words, one does invert each
path and change the origin to the end of the walk. An immediate consequence
for the complete ensemble of random paths is $\{\vec{u}_{i}\}=\{\vec{u}
_{i}^{\prime}\}$, with $i=1,2,...N$, which leads to $\{\vec{R}_{N}\}=\{\vec
{R}_{N}^{\prime}\}$. From the previous relations, it follows that $\{\vec
{R}_{N}\cdot\vec{u}_{N}\}=\{\vec{R}_{N}\cdot\vec{u}_{1}\}$, so the
configurational average $\langle\vec{R}_{N}\cdot\vec{u}_{N}\rangle=\langle
\vec{R}_{N}\cdot\vec{u}_{1}\rangle$ holds. This average, for $N\rightarrow
\infty$, is the persistence length $\lambda_{N}$. Therefore, the mean square
end-to-end distance could be rewritten as
\begin{equation}
\langle\vec{R}_{N}^{2}\rangle=\langle\vec{R}_{N-1}^{2}\rangle+2\lambda_{N}-1,
\label{disper_correla}%
\end{equation}
and we have established a relation between $\langle\vec{R}_{N}
^{2}\rangle$ and $\lambda_{N}$. We observed Eq.~\ref{disper_correla}
numerically, prior to its proof, by exact calculations for $N\leq24$.\ Some RW
models that obey such a relation are the $N$-step ensemble of ordinary RW and SAW paths.

\section{NUMERICAL RESULTS FOR THE SAW MODEL}
\label{sec:3}

From now on, we numerically study $\mathcal{I}_{j}$ for SAWs using the
non-reversed random walk (NRRW) algorithm to generate the ensemble of $N$-step non-overlapping paths. Because of the attrition problem, i.e., barriers or traps
that prevent paths to achieve $N$ steps, the NRRW is inefficient to generate good
statistics for long SAWs, since the probability decays as $p_{N}\propto\exp[-\gamma N]$, where $0<\gamma<1$ is the
attrition constant. However, the
generated data with this algorithm are surprisingly good enough to validate
our approach, showing that we choose the right corrections to scaling terms in
the expansion of IPLs.

\begin{figure}
[htpb]
\subfigure{\includegraphics[width=0.42\textwidth,height=0.28\textwidth]{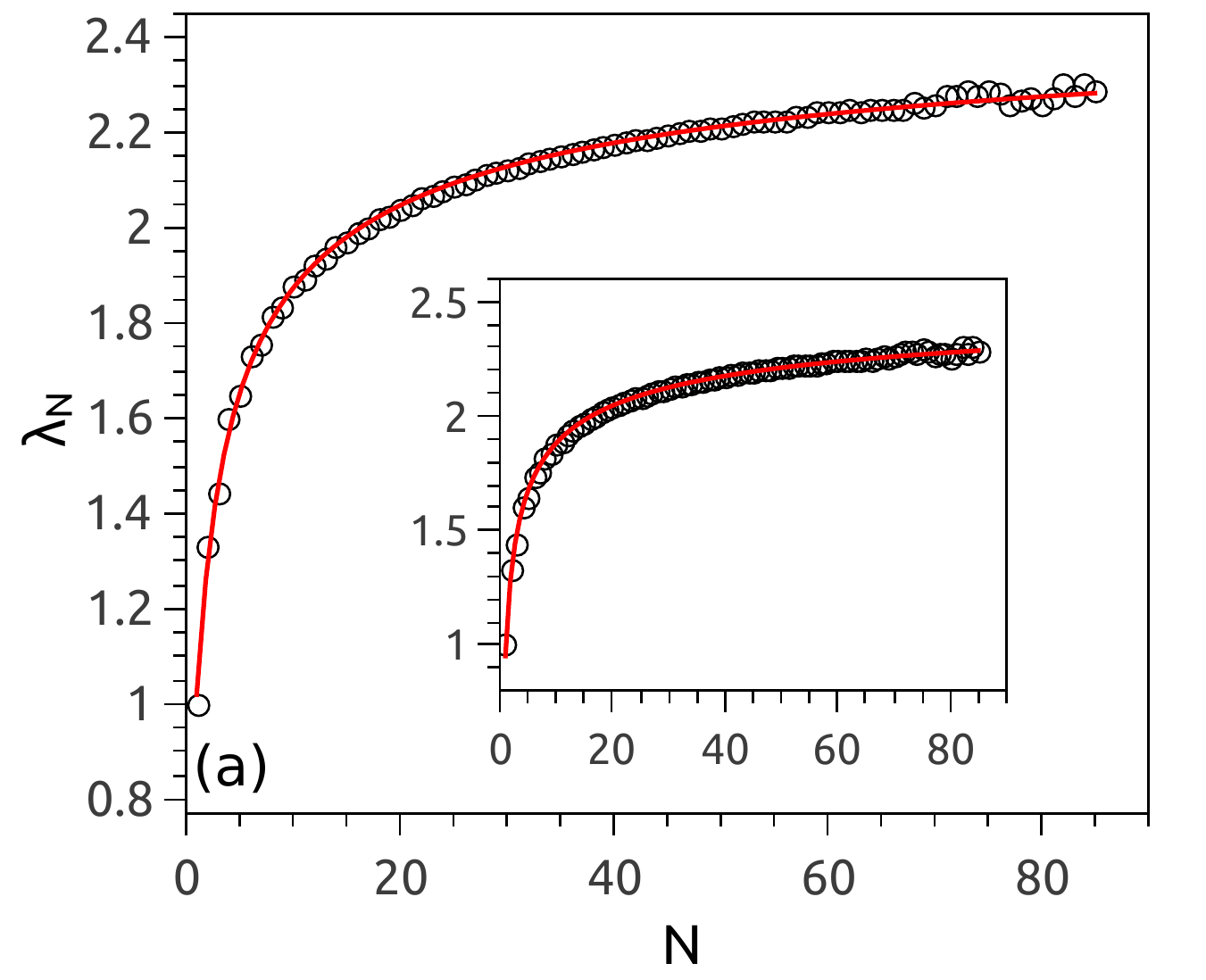}}
\subfigure{\includegraphics[width=0.42\textwidth,height=0.28\textwidth]{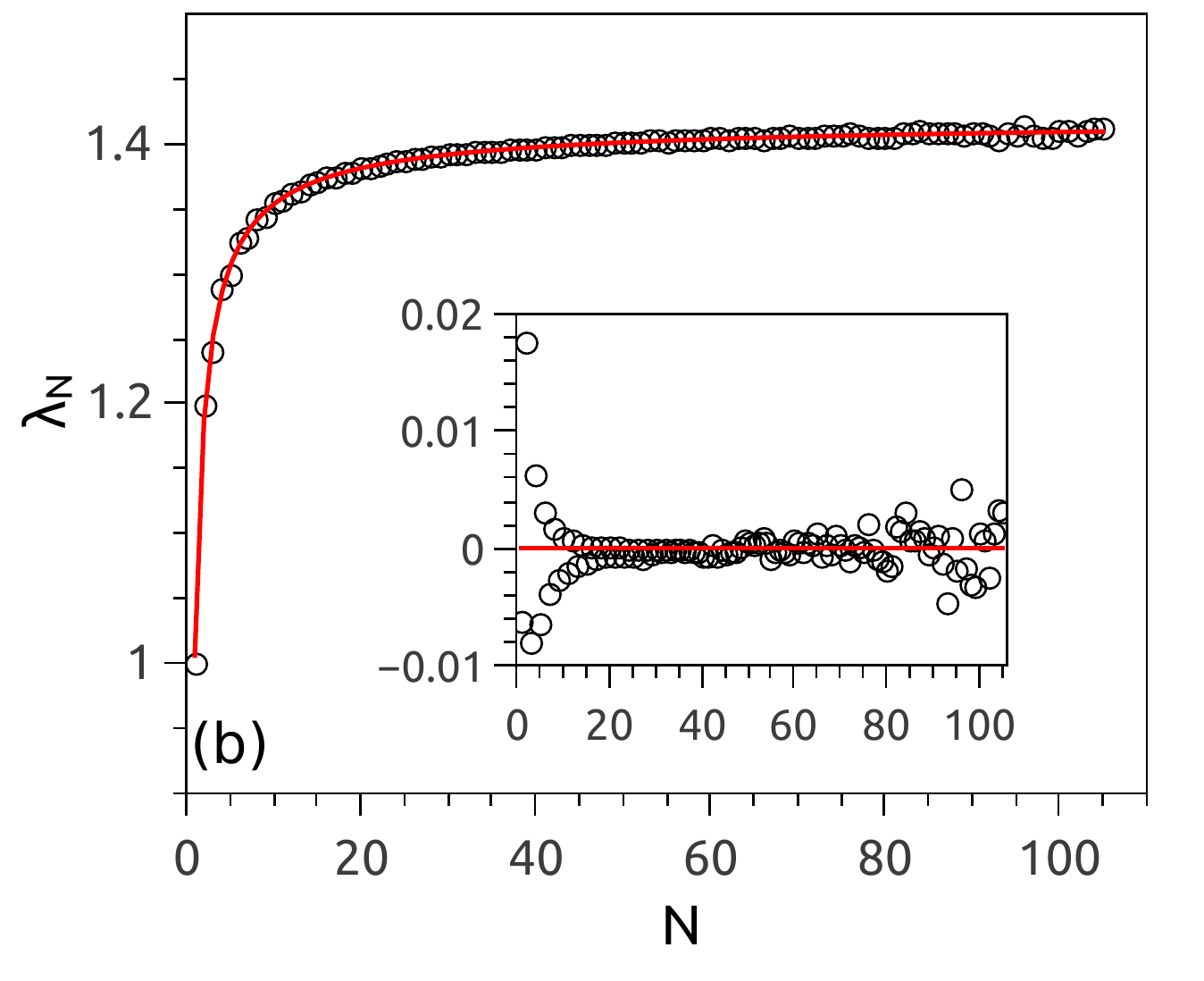}}
\caption{
SAW persistence length for (a) square and (b) cubic lattices. In both lattices, $\lambda_{N}$ converges to a constant.
The inset plot (a) shows $\lambda_{N}$ fitted by the function
$\alpha_{0}+\alpha_{1}N^{-0.34}$ of Ref.~\cite{JPA_Eisenberg_2003}, where $\lambda_{\infty}=2.664(3)$
is compatible with our estimate $\lambda_{\infty} = 2.525(4)$ from Eq.~\ref{persist_fit}. For the cubic lattice,
$\lambda_{\infty}\sim \sqrt{2}$ is compatible with the one of Ref.~\cite{AIP_Hagai_1990}. The inset plot (b) depicts the random pattern of the residual plot for $\lambda_{N}$ when fitted by Eq.~\ref{persist_fit}.
} \label{persistencia_plot}
\end{figure}

Starting with $\langle\vec{R}_{N}^{2}\rangle$, we now analyze the persistence
length. For the square lattice, $\nu_{0}=3/4$~\cite{JPA_Jensen_2004} and a common belief is that
$\Delta_{1}=3/2$~\cite{JSP_Caracciolo_2005raey}. With these exponents values, from Eq.~\ref{assintotic},
using only the first two leading exponents, we see that $\langle\vec{R}%
_{N}^{2}\rangle$ $\approx AN^{3/2}+BN^{1/2}$. The same reasoning leads to a
similar result for cubic lattices, where $\nu_{0}\sim0.587597(7)$ and
$\Delta_{1}\sim0.528(12)$ are widely accepted values \cite{PRL_Clisby_2010}. Both averages in Eq.~\ref{disper_correla}, $\langle\vec{R}_{N}^{2}\rangle$ and $\langle\vec{R}_{N-1}^{2}\rangle$, are obtained considering the same $N$-step ensemble. In this sense, we follow our previous notation by omitting the bracket index. The difference $\langle\vec{R}_{N}^{2}\rangle - \langle\vec{R}_{N-1}^{2}\rangle$ 
seems to be the discrete derivative of square end-to-end distance, which is not true for the SAW model. One should evaluate the derivative considering SAW ensembles of $N$ and $(N-1)$-steps: $\langle\vec{R}_{N}^{2}\rangle_{N} - \langle\vec{R}_{N-1}^{2}\rangle_{N-1}$. According to Eq.~\ref{assintotic}, the leading term of $\langle \vec{R}_{N}^{2}\rangle_{N}$ derivative is $N^{2\nu_{0}-1}$ with the first two corrections proportional to $N^{2\nu_{0}-2}$ and $N^{2\nu_{0}-\Delta_{1}-1}$, respectively. From the persistence length plots in Fig.~\ref{persistencia_plot}, $\lambda_{N}$ clearly does not diverge as the leading term of $\langle\vec{R}_{N}^{2}\rangle$ derivative, instead it seems to converge to a constant as $N$ goes to infinity~\cite{Note2}. Thus, we introduce the following series expansion:
\begin{equation}
\lambda_{N}=\alpha_{0}+\alpha_{1}N^{-w_{1}}+\alpha_{2}N^{-w_{2}}
+\cdots\label{persist_fit}
\end{equation}
where the exponents $w_{i}>0$, $i=1,2,3\cdots$, are linear combinations of
$\nu_{0}$ with analytical and non-analytical corrections to scaling exponents.
As for example, from the persistence length data fitting with
Eq.~\ref{persist_fit} (see Fig.~\ref{persistencia_plot}), we find that
$w_{1}=2\nu_{0}-2$ and $w_{2}=2\nu_{0}-\Delta_{1}-1$ are the best choices. The
$\alpha_{i}$ and $w_{i}$ values are shown in Tab.~\ref{tab:const}. An
immediate consequence of such findings along with Eq.~\ref{disper_correla}, is
that $\lambda_{N}$ could contribute only with second and higher-order of
analytic and non-analytic corrections for $\langle\vec{R}_{N}^{2}\rangle$. Our
estimate of $\lambda_{N}$, for square lattices, is compatible with the one of
Eisenberg and Baram \cite{JPA_Eisenberg_2003}. Through their estimate of the
step-step correlation scaling: $\langle\vec{u}_{1}\cdot\vec{u}_{j}%
\rangle=\langle\xi_{1,j}\rangle_{N}\sim0.6j^{-1.34(5)}$, and the definition
$\lambda_{N}=\sum_{j=1}^{N}\langle\xi_{1,j}\rangle_{N}$, we obtained
$\lambda_{N}\sim\alpha_{0}-1.7N^{-0.34(5)}$, with which we fitted the
persistence length data, but leaving $\alpha_{1}$ free, as shown in the inset
of Fig.~\ref{persistencia_plot}(a).

\begin{table}
\caption{\label{tab:const} Coefficients and exponents for fitting, with Eq.~\ref{persist_fit}, the $\lambda_{N}$ data obtained from simulations for 2D and 3D
square lattices. The $w_{1}=0.34(5)$ value is an effective exponent, thus depending on the coefficients $\alpha_{i}$
and exponents $w_{i}$ of
Eq.~\ref{persist_fit}~\cite{JSP_Caracciolo_2005raey}.} \begin{ruledtabular}
\begin{tabular}{cccccc}
$d$ &  $\alpha_{0}$ & $\alpha_{1}$ & $\alpha_{2}$ & $w_{1}$ & $w_{2}$ \\ \hline
$2$ &  $2.525(4)$ & $-2.32(3)$ & $0.81(3)$ & $0.5$ & $1$         \\
$2$\footnote{Fitting with equation $\lambda_{N}\sim\alpha_{0}+\alpha_{1}N^{-0.34(5)}$ from
Ref.~\cite{JPA_Eisenberg_2003}.} &  $2.664(3)$ & $-1.714(9)$ & $-$ & $0.34$ & $-$  \\
$3$ &  $1.422(1)$ & $-0.39(6)$ & $-0.022(5)$  & $0.8248$ & $0.34$ \\
\end{tabular}
\end{ruledtabular}
\end{table}
\begin{figure}
[htpb]\centering
\subfigure{\includegraphics[width=0.42\textwidth,height=0.28\textwidth]{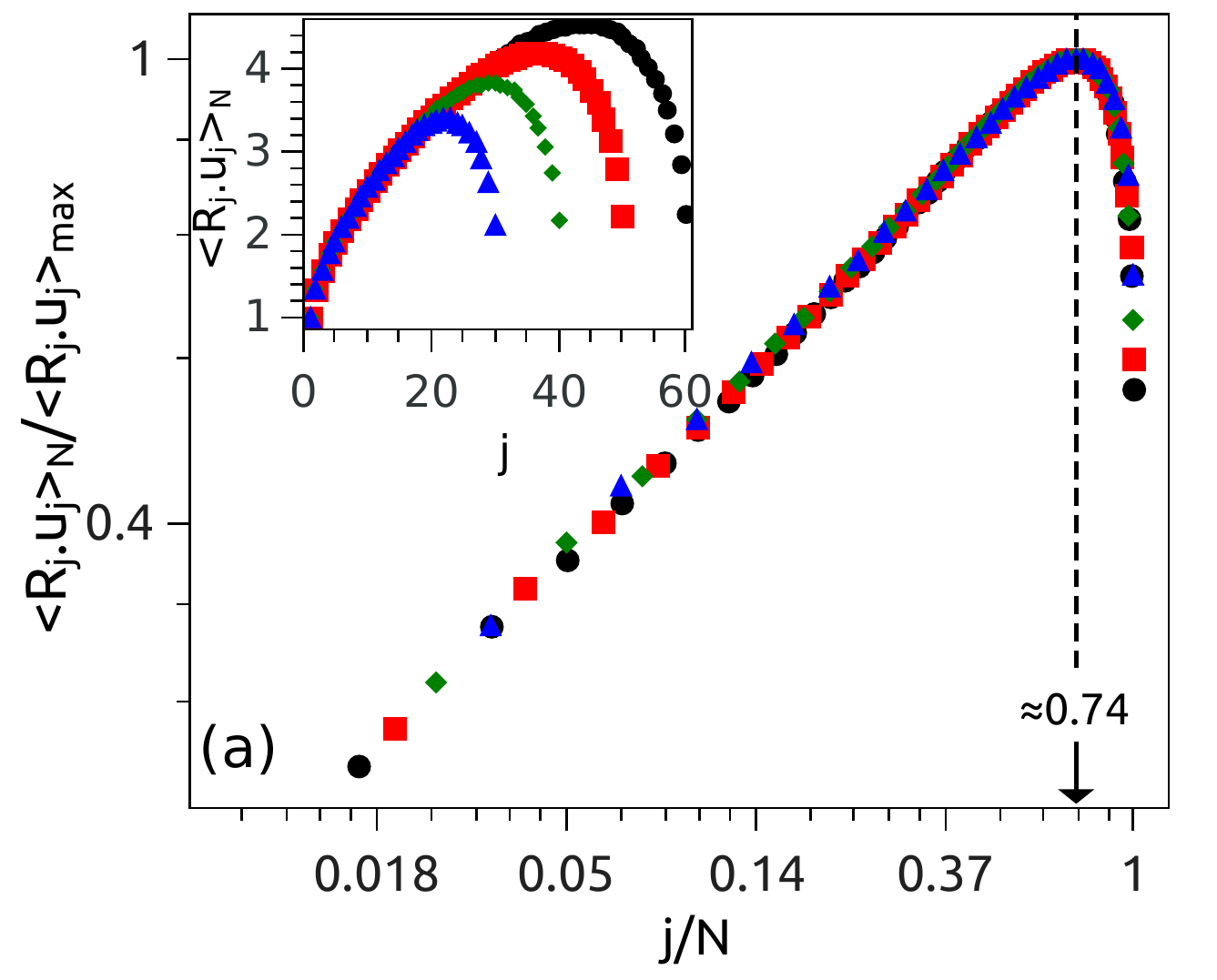}}
\subfigure{\includegraphics[width=0.42\textwidth,height=0.28\textwidth]{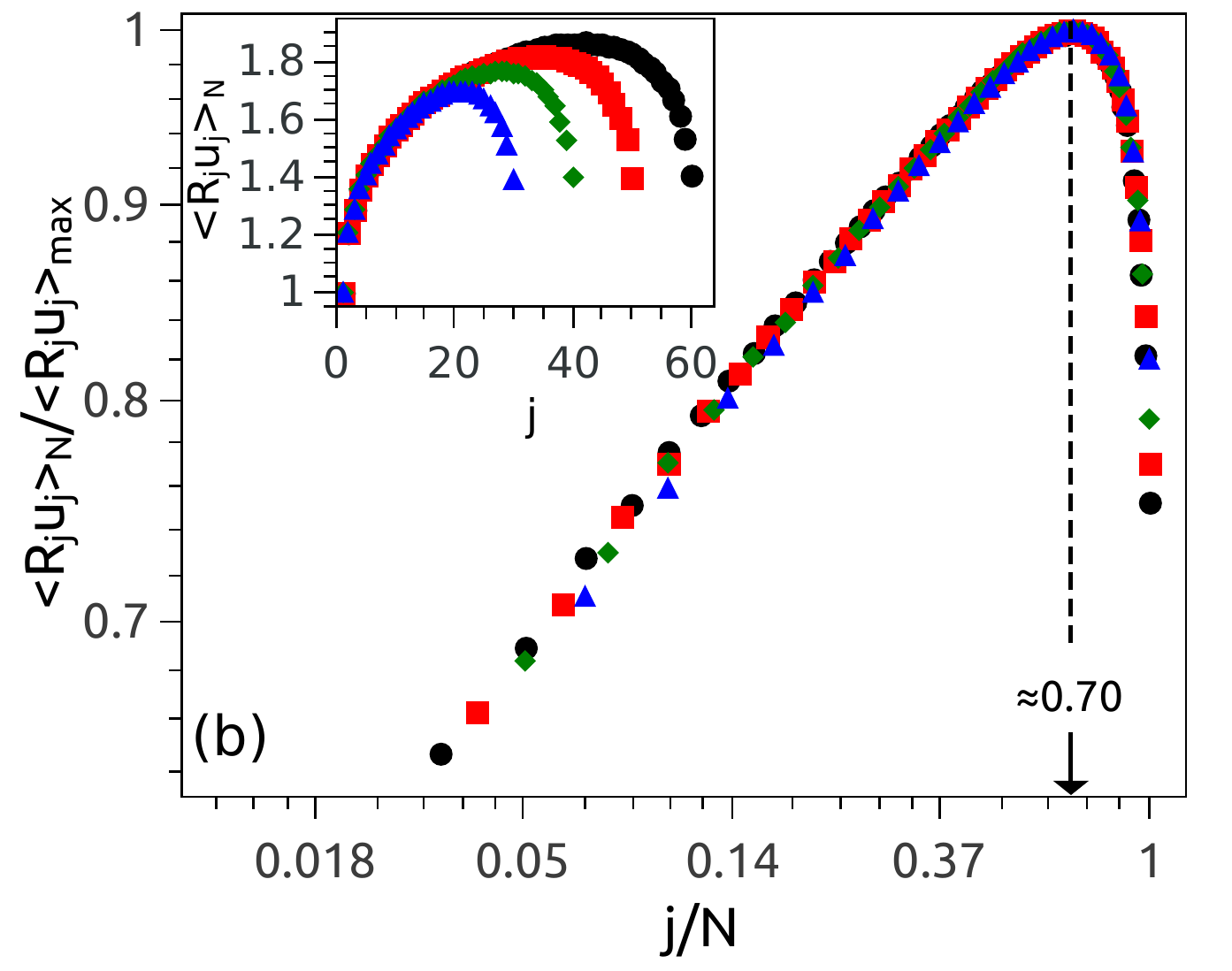}}
\caption{
IPL data collapse in $\log$-$\log$ scale for (a) square and (b) cubic lattices with ($\begingroup\color{blue}\blacktriangle\endgroup\,N=30$), ($\begingroup\color{green!60!black}\blacklozenge\endgroup\,N=40$), ($\begingroup\color{red}\blacksquare\endgroup\,N=50$) and ($\bullet\,N=60$).
The $\langle \vec{R}_{j} \cdot \vec{u}_{j}\rangle_{N}$ behaves as linear increasing function up to $\sim j_{max}$, with a slope $\approx (2\nu_{0}-1)$. In both lattices $j_{max}\propto N$, with constant of proportionality close to each other ($\sim 0.7$). For $j>j_{max}$ the scalar products contribute to residual terms of corrections to scaling of $\langle\vec{R}_{N}^{2}\rangle$.
} \label{dot_inter_fitting}
\end{figure}

\begin{figure}
[htpb]
\subfigure{\includegraphics[width=0.42\textwidth,height=0.28\textwidth]{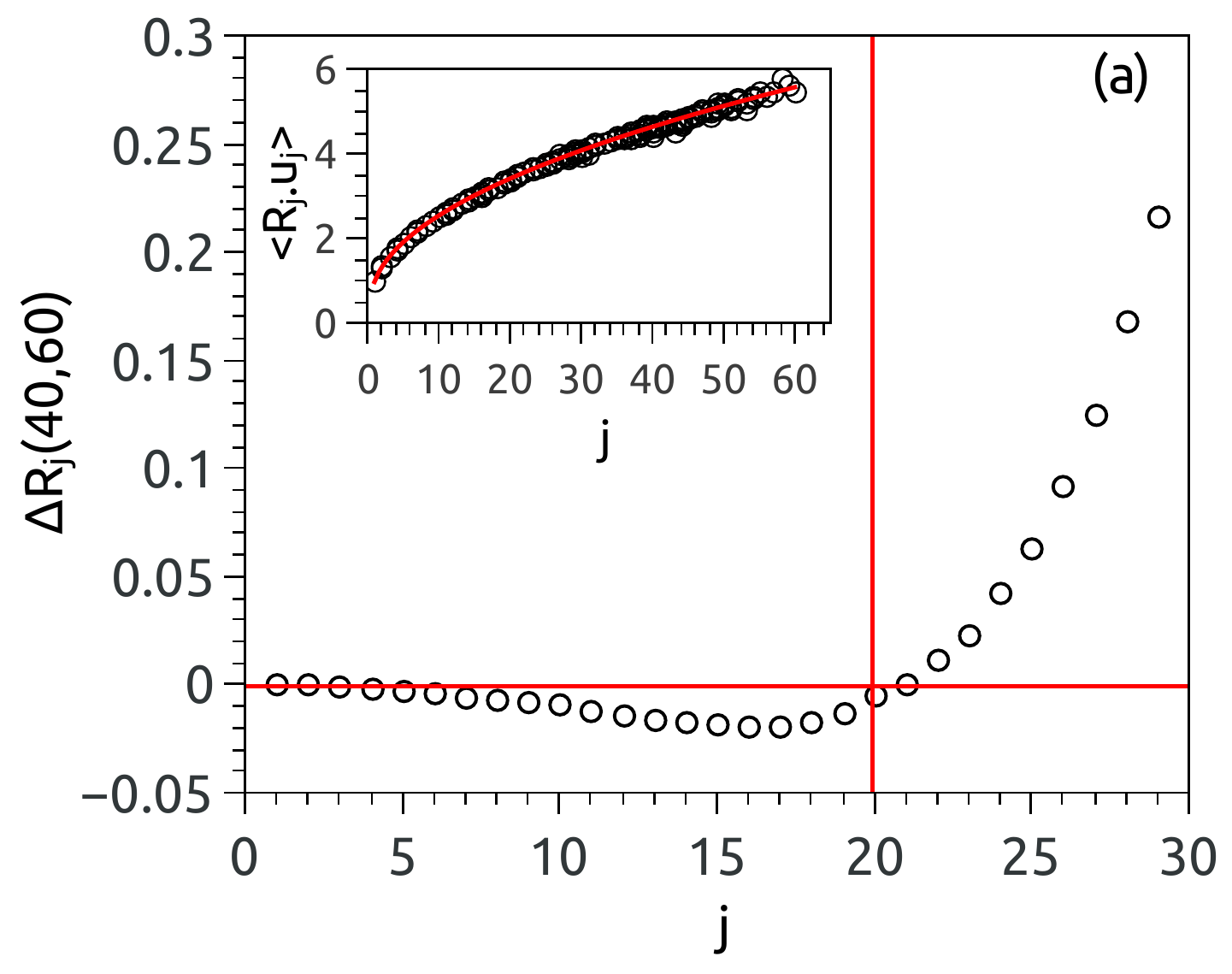}}
\subfigure{\includegraphics[width=0.42\textwidth,height=0.28\textwidth]{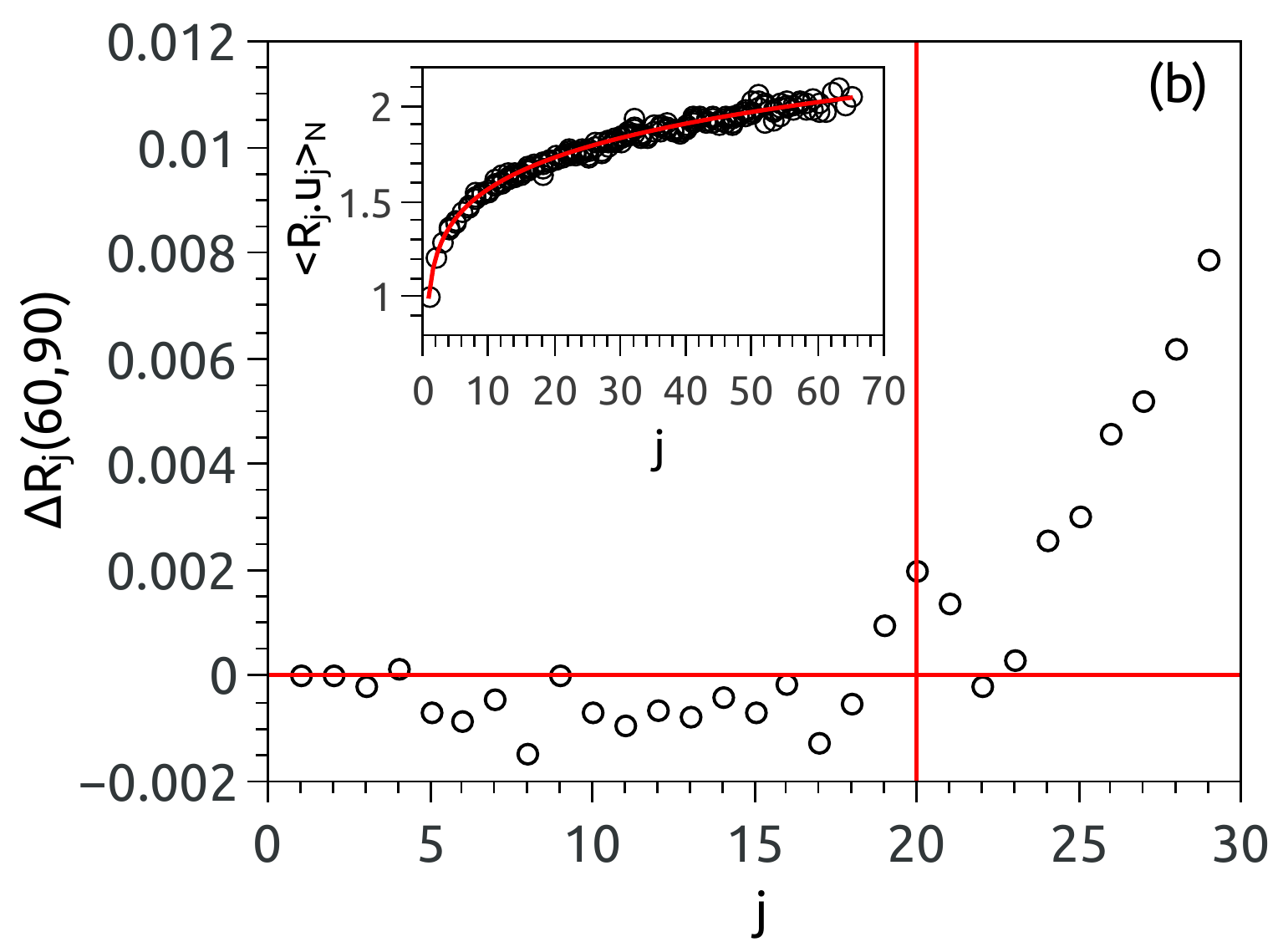}}
\caption{
IPL differences ($\Delta R_{j}(N_{1},N_{2})$) for SAWs: (a) for square and (b) cubic lattices, with $N_{1}=40$ and $N_{2}=60$,
and $N_{1}=60$ and $N_{2}=90$, respectively.
According to $\Delta R_{j}(N_{1},N_{2})$ depicted here [see Eq.\ref{diff_ensemble}], the $\mathcal{I}_{j}$ starts to be influenced for $j_{c}>N_{1}/2$ and $j_{c}>N_{1}/3$, for 2D and 3D square lattices, respectively.
Inset plots show IPL non-weighted fit using Eq.~\ref{scaling:sip}, within a confidence interval of $95\%$.
The square lattice data includes $N$ ranging from $90$ to $120$ with increment of 10. The fitting parameters obtained are $\varphi=0.4979(21)$
and $\beta_{1}=0.6630(50)$. The cubic lattice data includes $N$ ranging from $150$ to $195$ with increment of 15. The fitting
parameters obtained are $\varphi=0.1752(14)$, $\Delta_{1}=0.522(52)$, and $\beta_{1}=0.7581(56)$.}
\label{diff_prod_inter}
\end{figure}

Now, consider $\mathcal{I}_{j}$, for $1<j<N$. According to the collapsed
$\log\times\log$ plots of Fig.~\ref{dot_inter_fitting}, it is notable that
$\mathcal{I}_{j}$ looks like a straight line up to near the point where it
reaches its maximum value, at the $j_{max}$ step, with a positive slope
$\approx$ $2\nu_{0}-1$. From Eqs.~\ref{assintotic} and~\ref{observ_quad2}, and
Fig.~\ref{dot_inter_fitting}, assuming $\mathcal{I}_{j}$ scales as
$j^{2\nu_{0}-1}$ is reasonable, at least for $j<j_{max}$. Such proportionality
leads us to look for reliable estimates of $\nu_{0}$, and corrections to
scaling exponents, for SAW ensembles with $N$ not too large. To accomplish
this aim, diminishing the influence of the $N$-step ensemble on estimates of
scaling exponents is necessary. In other words, it is necessary to find a cutoff
step $j=j_{c}(N)$, at which $\mathcal{I}_{j}$ begins to be noticeably
influenced by the $N$-step SAW ensemble. Surely, we can neglect steps above
$j_{max}$. To seek the $j_{c}(N)$ step, we use the difference between the IPLs
of two $N$-step ensembles, one that contains $N_{1}$, and the other $N_{2}$
steps,
\begin{equation}
\Delta R_{j}(N_{1},N_{2})=\langle\vec{R}_{j}\cdot\vec{u}_{j}\rangle_{N_{2}%
}-\langle\vec{R}_{j}\cdot\vec{u}_{j}\rangle_{N_{1}}, \label{diff_ensemble}%
\end{equation}
where $N_{2}>N_{1}$. According to Fig.~\ref{diff_prod_inter}(a), the IPL has
approximately the same behavior for the two path lengths, up to the middle of
the shortest path, $j_{c}(N_{1})\sim N_{1}/2$, for square lattices. Similarly,
for cubic lattices, it has the same behavior, up to a third of the shortest
path $j_{c}(N_{1})\sim N_{1}/3$ [see Fig.~\ref{diff_prod_inter}(b)].
Therefore, using $j\leq j_{c}(N)$, with $j_{c}(N)=N/2$ and $j_{c}(N)=N/3$ for
2D and 3D lattices, respectively, it is suitable to estimate the scaling
exponents through $\mathcal{I}_{j}$.

Additional information to do scaling analysis with $\mathcal{I}_{j}$ comes
from the expansion of $\langle\vec{R}_{N}^{2}\rangle$ in powers of $N$. We
have found no evidence of the linear term in the expansion of $\langle\vec
{R}_{N}^{2}\rangle$ on square or cubic lattices. The nonexistence of the
linear term is also reported in Refs.~\cite{JPA_Chan_2012,JPA_Lothar_1999}.
From Eq. \ref{observ_quad2}, the only way to disappear with the linear term in
the expansion of $\langle\vec{R}_{N}^{2}\rangle$ is if the summation of
$\mathcal{I}_{j}$ cancels it out. This finding, with Eq. \ref{assintotic},
leads us to write
\begin{equation}
\mathcal{I}_{j}=\beta_{0}+\beta_{1}(j-\tau)^{\varphi}\left[  1+\beta
_{2}(j-\tau)^{-\Delta_{1}}+...\right]  , \label{scaling:sip}%
\end{equation}
for $j\leq j_{c}(N)$, where $\tau$ is a smoothing constant
\cite{JPA_Clisby_2007}. We set $\beta_{0}=1/2$ just to cancel the linear term.
Also, we did another ansatz: $\beta_{2}=-\left(  2\nu_{0}-1\right)  $ and
$\tau=0.5$. This was inspired by results considering only the first
non-analytical correction to scaling term, and leaving only the parameters
$\beta_{1}$, $\beta_{2}$ and $\tau$ free, which lead us to find $\beta
_{2}\approx-\left(  2\nu_{0}-1\right)  $ for the 3D case. Notice that, in
general $\beta_{1}=\beta_{1}\left(  N\right)  $ and $\beta_{2}=\beta
_{2}\left(  N\right)  $; however, for $N$ not too large, order of hundreds for
2D and 3D cases, these parameters converged to constants, for $j\leq j_{c}(N)$.

The IPL data, containing several $N$-step ensembles, fitted by
Eq.~\ref{scaling:sip} is depicted in the inset plots of
Fig.~\ref{diff_prod_inter}. For both, the 2D and 3D square lattices, the
leading and sub-leading exponents are in excellent agreement with the believed
results. For the square lattice, we found $\nu_{0}=0.7489(21)$, and the
non-analytical first exponent results in $\Delta_{1}=3/2$; because it does not
appear in Eq.~\ref{scaling:sip}, showing that there exists a constant in the
expansion of $\langle\vec{R}_{N}^{2}\rangle$. This is confirmed through the
expansion of $\lambda_{N}$; the predicted results are $\varphi=0.5$ and
$\Delta_{1}=3/2$. For cubic lattices we found $\nu_{0}=0.58757(140)$, and
$\Delta_{1}=0.522(52)$, while the best predicted results are $\nu
_{0}=0.587597(7)$ and $\Delta_{1}=0.528(12)$ \cite{PRL_Clisby_2010}. Using
several $N$-step ensembles seeks to reduce the error on exponent estimates;
however, they may carry some small biased errors. To check this, for 2D-SAW,
we used $N=120$ steps obtaining $\nu_{0}=0.7500(63)$, and for 3D-SAW we used
$N=198$ steps giving $\nu_{0}=0.58758(450)$ and $\Delta_{1}=0.52(17)$.
However, the errors we get are not as small as those from literature for the
3D case \cite{PRL_Clisby_2010}. We can improve these results, by taking into
account the advantage of the statistical invariance, and calculating the IPL
starting from the end of the generated chains, thus doubling the sample. In
fact, it is out of the scope of this paper to find high precision values for
the exponents, but to validate and evaluate the benefits of our approach.
Moreover, the whole potential of the method to do the scaling analysis of RWs
has not been fully exploited. We expect that the corrections to scaling
exponents are easily accessible from the study of the monotonically decreasing
$\mathcal{I}_{j}$ terms of $\langle\vec{R}_{N}^{2}\rangle$, which will readily
be tackled.

\section{CONCLUDING REMARKS}
\label{sec:4}

In summary, we have proposed an approach to address the scaling of RW
conformational quantities, where the mean square end-to-end distance is
proportional to the summation of the inner persistence length, $\mathcal{I}
_{j}=\langle\vec{R}_{j}\cdot\vec{u}_{j}\rangle$. For RW models, where paths
obtained by orthogonal transformations occur with the same probability, we
obtained a novel relation between the mean square end-to-end distance and
persistence length. Despite the numerical limitations to do scaling analysis,
we introduce a series for the persistence length $\lambda_{N}$ and show that
it converges to a constant, $\alpha_{0}$, apart corrections to scaling terms.
We also developed a method to calculate the scaling exponents from
$\mathcal{I}_{j}$ with a path cutoff that diminishes the $N$-step ensemble
influence. Thus, the method is efficient to obtain the scaling behavior of SAW. 

We conclude that only an ensemble of paths with the same length is
sufficient for performing scaling analysis, being that the whole information
needed are contained in the inner part of the paths. The scaling method
discussed in this paper can be important for studying universality,
criticality, and conformational properties of systems mapped on RW models, such as polymers,
biopolymers, and magnetic systems.

\section{ACKNOWLEDGMENTS}

The authors thank T. J. Arruda, A. Caliri, J.~C. Cressoni, G.~M.
Nakamura, and F.~L. Ribeiro for their valuable comments. Also, the authors
thank M. V. A. da Silva for exact enumeration SAW code and C. Traina for
helpful discussions. C.~R.~F. Granzotti acknowledges CAPES for financial
support. A.~S. Martinez acknowledges CNPq (Grants No. 400162/2014-8 and No. 307948/2014-5) and
NAP-FisMed for support. M.~A.~A. da Silva thanks Professor R. H. Swendsen for
early fruitful discussions about scaling theory and SAWs and also,
acknowledges FAPESP (Grants No. 11/06757-0 and No. 12/03823-5) for financial support.


\end{document}